\documentclass[aps,prl,twocolumn,amsmath,amssymb,floats]{revtex4-1}
\usepackage{graphicx}
\usepackage{amsmath}
\usepackage{epstopdf}
\usepackage{amsbsy}
\usepackage{color}
\usepackage{dcolumn}
\usepackage{bm}
\usepackage{mathtools} 

\begin{document}

\title{Enhancing Superconductivity by Disorder}
\author{Maria N. Gastiasoro}
\altaffiliation{Current address: School of Physics and Astronomy, University of Minnesota, Minneapolis, MN 55455, USA.}
\author{Brian M. Andersen}
\affiliation{Niels Bohr Institute, University of Copenhagen, Juliane Maries Vej 30, 2100 Copenhagen,
Denmark}

\date{\today}

\begin{abstract}

We study two mechanisms for enhancing the superconducting mean-field transition temperature $T_c$ by nonmagnetic disorder in both conventional (sign-preserving gaps) and unconventional (sign-changing gaps) superconductors (SC). In the first scenario, relevant to multi-band systems of both conventional and unconventional SC, we demonstrate how favorable density of states enhancements driven by resonant states in off-Fermi-level bands, lead to significant enhancements of $T_c$ in the condensate formed by the near-Fermi-level bands. The second scenario focuses on systems close to localization where random disorder-generated local density of states modulations cause a boosted $T_c$ even for conventional single-band SC. We analyze the basic physics of both mechanisms within simplified models, and discuss the relevance to existing materials. 
\end{abstract}

\maketitle

{\it Introduction.} What happens to the superconducting (SC) transition temperature $T_c$ upon increasing the amount of disorder in a material? This important question has been thoroughly studied both experimentally and theoretically, and the answer is known to depend on the nature of the disorder and the pairing symmetry of the SC. The naive answer, in overall agreement with the bulk of previous studies, is that $T_c$ drops, or remains unaffected, at best. The latter possibility is the essence of Anderson's theorem stating that nonmagnetic disorder does not affect $T_c$ for conventional SC.\cite{anderson} This ceases to be true for unconventional SC with sign-changing gap functions, and attention has been centered on measuring and explaining the $T_c$ suppression rate, i.e $dT_c/dx$ where $x$ denotes the concentration of disorder.\cite{balatskyrmp} 

There is, however, no fundamental principle preventing $T_c$ from rising with increased disorder, and experimental reports of disorder-enhanced $T_c$ exist in the literature.\cite{matthias59,Slebarski,hammerath10,kikoin,teknowijoyo} It is also possible that part of the $T_c$ value in doped systems, such as cuprates and iron-pnictides, or inhomogeneous\cite{cava} or granular SC\cite{konig} arise from the inhomogeneity itself. This idea is in line with a number of earlier theoretical studies concluding that in conventional SC disorder may under some circumstances enhance $T_c$.\cite{Feigelman,Feigelman10,Burmistrov,Mayoh,Yukalov1,Yukalov2,Mayoh2,Palestini} For example, in systems with short-range (screened) Coulomb interactions, $T_c$ may be strongly enhanced by Anderson localization, a property related to the multifractality of the wavefunctions in the disordered system.\cite{Feigelman,Burmistrov,Mayoh} Another series of studies have focussed on periodically modulated SC, and found that such systems may also exhibit larger $T_c$ than the homogeneous case.\cite{martin,tsai,maier,barush,goren,mondaini} These results raise the general question; under what circumstances does inhomogeneity boost $T_c$? Pinpointing such conditions may guide new disorder-engineered SC with elevated  $T_c$.

\begin{figure}[t]
\begin{center}
 \includegraphics[width=\columnwidth]{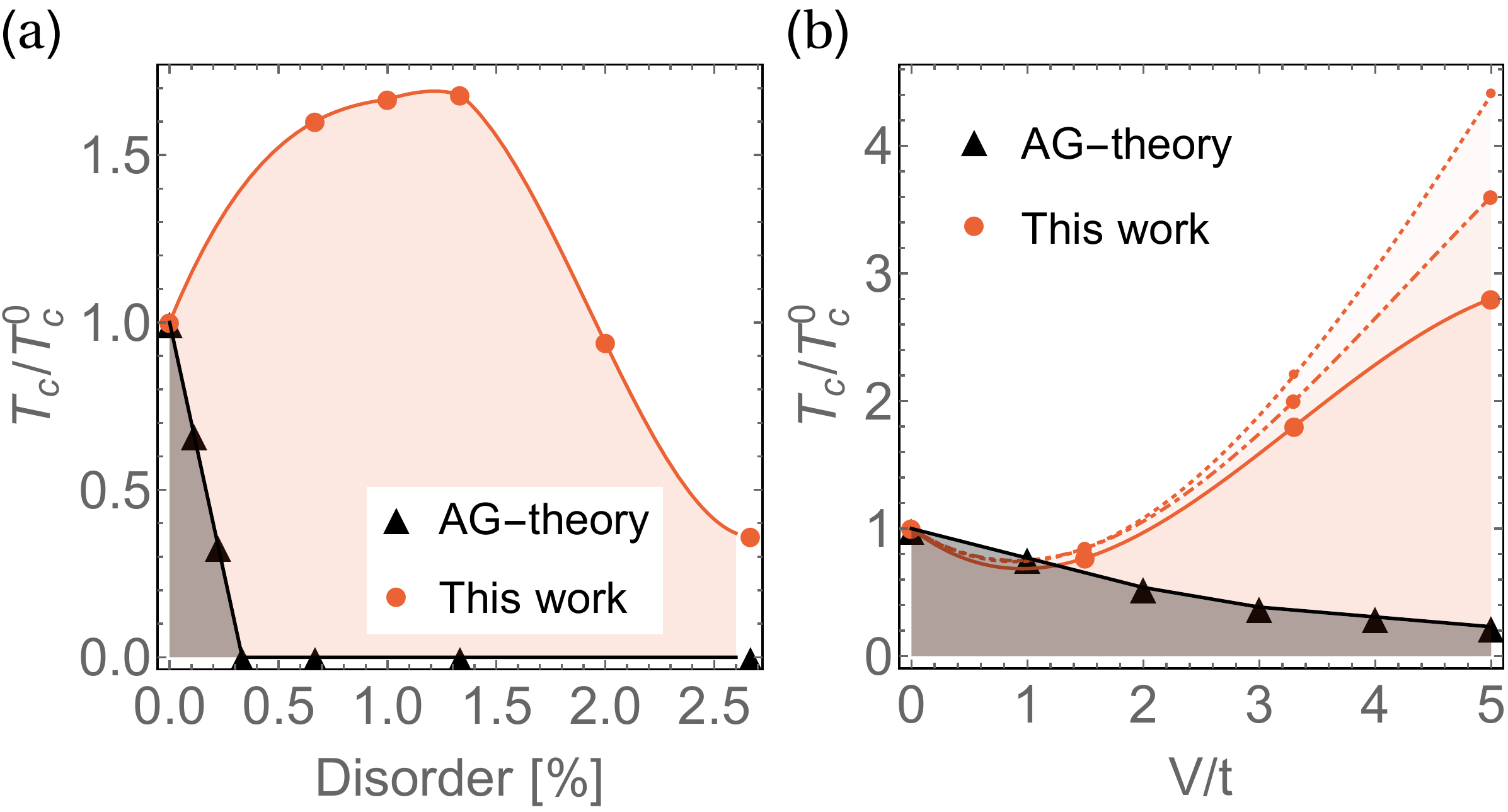}
\end{center}
\caption{Superconducting critical temperature $T_c/T_c^0$ in the presence of nonmagnetic disorder in (a) an unconventional multi-band $s_{\pm}$ SC versus disorder concentration, and (b) a one-band $s$-wave SC with $15\%$ disorder versus impurity strength $V$. The black curves show the results when disallowing spatial modulations of density and SC OP consistent with AG-theory. The red curves show the self-consistent cases with spatial modulations of both quantities. In (b) $p=2\%$ (dotted), $p=5\%$ (line-dotted) and $p=10\%$ (solid), see text.}
\label{fig:1}
\end{figure}

In this paper, we demonstrate that disorder-generated $T_c$-enhancements can happen for both conventional and unconventional SC. Our study highlights the crucial role of spatial inhomogeneity and the generation of favorable local density of states (LDOS) enhancements generated by nonmagnetic disorder. This goes beyond the standard Abrikosov-Gor'kov (AG) treatment of disordered SC assuming spatially uniform SC order parameter (OP) and constant DOS.\cite{AG1}  We study two separate scenarios for disorder-generated $T_c$-enhancements: 1) dilute disorder in multi-band SC, and 2) dense disorder in conventional one-band SC. In the former case 1), the multi-band property is crucial; impurity resonant states generated by non-SC off-Fermi-level bands generate LDOS enhancements at the Fermi level $E_F$, which, through interband coupling, feeds into the near-Fermi-level bands important for SC. As seen from Fig.~\ref{fig:1}(a), even for unconventional SC this effect can overwhelm pair-breaking caused by nonmagnetic disorder, and raise $T_c$ well above that of the homogeneous system, $T_c^0$.  In the second case 2), the band structure is unimportant; LDOS modulations allow for regions with increased DOS in a densely disordered normal state which can lead to enhanced $T_c$.\cite{Feigelman,Burmistrov,Mayoh} Normally the insensitivity of the OP to disorder in conventional SC, is understood with reference to Anderson's theorem.\cite{anderson} This, however, relies on dilute disorder and spatially uniform OP and DOS. We do not include the harmful effect of longer-range Coulomb repulsion,\cite{Maekawa,Anderson83,Coffey,Finkelstein} restricting the relevance to sufficiently screened systems.\cite{Burmistrov} We also stress that our studies refer to the mean-field $T_c$, and that the role of phase fluctuations remain an important outstanding question.

To the best of our knowledge, scenario 1) has not been pointed out before, and for 2) even though superconductivity near the localization threshold has been discussed before\cite{Feigelman10,Ma85}, only a limited set of previous studies have discussed the {\it favorable} effects of nonmagnetic disorder in conventional BCS superconductors.\cite{Feigelman,Burmistrov,Mayoh,zhitomirsky,semenikhin} In particular, Burmistrov {\it et al.}\cite{Burmistrov} studied the interplay of disorder and SC within an RG-analysis of the nonlinear $\sigma$-model, inferring that Anderson localization enhances $T_c$ for both 2D and 3D systems. Notably this enhancement effect was not, however, observed in earlier numerical finite size system simulations of the disordered attractive Hubbard model.\cite{trivedi1996,ghosal_phase,bouadim}

For unconventional SC, the importance of allowing for spatial inhomogeneity in the SC OP has been pointed out for cuprates and heavy fermion SC.\cite{zhitomirsky,semenikhin,franz,semenikhin_d,ghosal_d,das11} In the case of cuprates, the observed $T_c$-suppression rate is considerably weaker than dictated by AG-theory,\cite{exp_cuprates1,exp_cuprates2,exp_cuprates3,exp_cuprates4,exp_cuprates5,exp_cuprates6} which was ascribed to the importance of a spatially adaptive SC condensate.\cite{zhitomirsky,franz,semenikhin_d,ghosal_d} We note that the enhancement of SC by disorder from the perspective of local enhanced pairing interaction has been also discussed in the literature.\cite{nunner,andersen,maska,kemper,foyevtsova,astrid,astridny} Within this scenario, the pairing interaction itself gets locally enhanced by disorder. Finally we note that $T_c$-enhancements from disorder has been also discussed in the context of negative-$U$ centers.\cite{negU1,negU2,negU3,negU4}

{\it Scenario 1.} For concreteness we demonstrate the $T_c$-enhancement mechanism by a multi-band model relevant to unconventional iron-based SC (FeSC)
\begin{equation}\label{eqH}
\mathcal{H} = \mathcal{H}_0 + \mathcal{H}_{BCS} + \mathcal{H}_{imp},
\end{equation}
where $\mathcal{H}_0=\sum_{\mu\nu\sigma ij} (t_{ij}^{\mu\nu} - \mu_0 \delta_{ij} \delta_{\mu\nu})\hat{c}^\dagger_{i\mu\sigma} \hat{c}_{j\nu\sigma}$ denotes the hopping Hamiltonian with parameters adapted from the five-band model of Ref. \onlinecite{ikeda}. The band structure consists of a Fermi surface with both electron and hole sheets with orbital $t_{2g}$ character, and lower lying bands some of which exhibit predominantly $e_g$ character.\cite{ikeda} The operator $\hat{c}^\dagger_{i\mu\sigma}$ creates an electron at site $i$ in orbital state $\mu$ with spin $\sigma$, and $\mu_0$ is the chemical potential adjusting the average electron density $n$ of 6.0 electrons per site. We stress that the disorder does not provide additional carriers. The indices $\mu$ and $\nu$ denote the five iron orbitals ($d_{xz}$, $d_{yz}$, $d_{xy}$, $d_{x^2-y^2}$, $d_{3z^2-r^2}$). Superconductivity is included through the standard multi-orbital singlet pairing BCS term, $\mathcal{H}_{BCS}=\sum_{i\neq j,\mu\nu} [\Delta_{\mu\nu} \hat{c}^\dagger_{i\mu\uparrow} \hat{c}^\dagger_{j\nu\downarrow} + \mbox{H.c.}]$. We fix the SC coupling constant $\Gamma=0.2$ eV for attractive next-nearest neighbor (NNN) intra-orbital (and orbital independent) pairing, producing a sign-changing $s\pm$ SC ground state with $\Delta_{\mu}^0=(0.78,0.78,1.31,0.063,0.055)$ meV and $T_c^0=21$ K in the homogeneous case.\cite{gastiasoro16} We will be interested in 3D materials like cuprates and FeSCs which are layered quasi-2D systems. Therefore we can perform computationally simpler 2D calculations, with the understanding, however, that it is the inter-planar coupling that supports a finite $T_c$. Finally $\mathcal{H}_{imp} = \sum_{\mu\sigma\{i\}} V \hat{c}^\dagger_{i\mu\sigma} \hat{c}_{i\mu\sigma}$ is the impurity term consisting of a set of impurity sites $\{i\}$ with onsite potential $V$ assumed, for simplicity, to be orbital independent and of intra-orbital nature, in overall agreement with DFT findings.\cite{nakamura,kreisel} We solve Eq.~(\ref{eqH}) by finding self-consistent solutions to its corresponding Bogoliubov-de Gennes (BdG) equations on $30\times 30$ lattices with unrestricted density and OP fields with respect to all orbital and site degrees of freedom. For further computational details we refer to the supplementary material,\cite{SM} and our earlier publications.\cite{gastiasoro13,gastiasoro152,gastiasoro16}

\begin{figure}[t]
\begin{center}
 \includegraphics[width=\columnwidth]{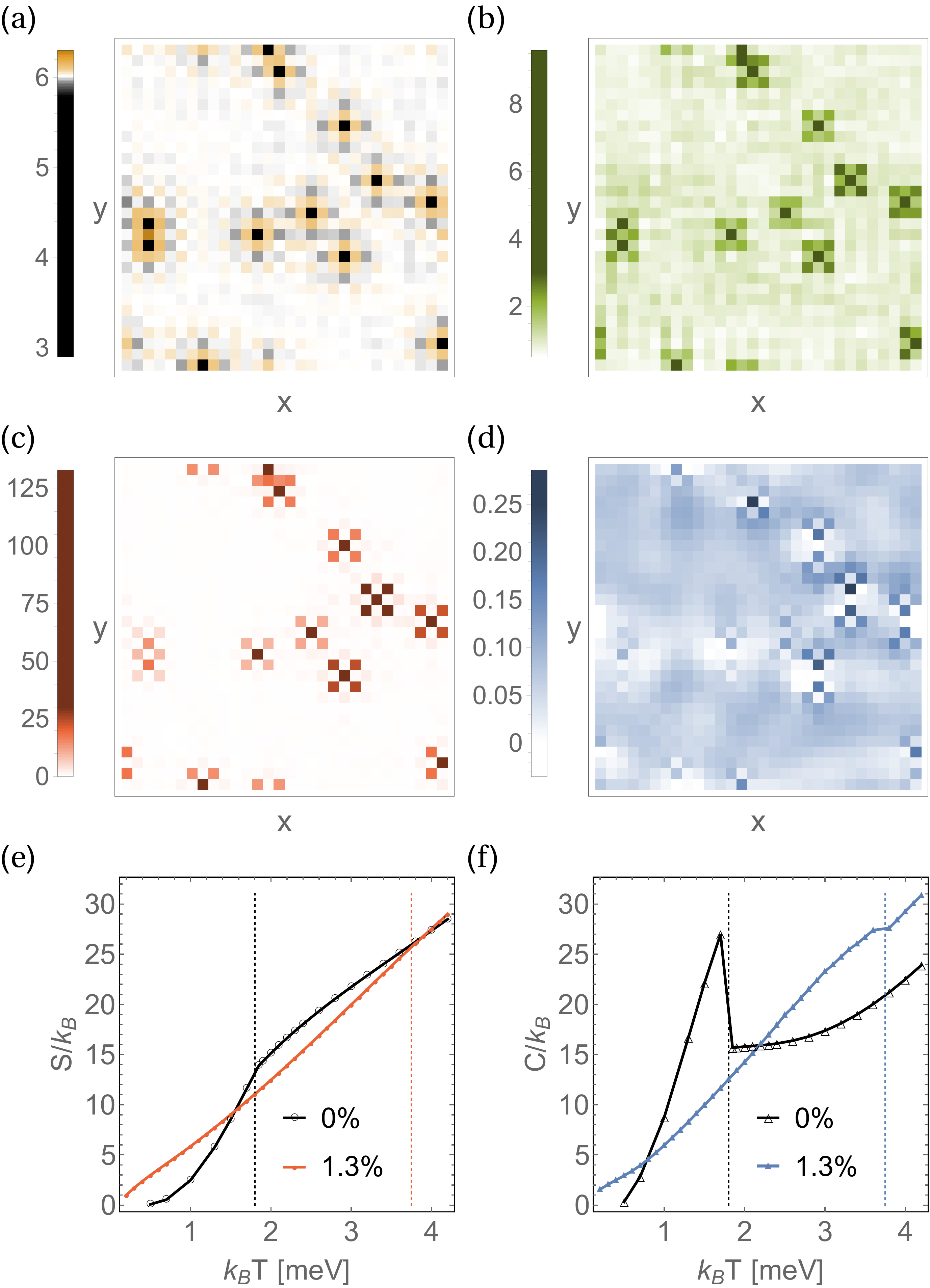}
\end{center}
\caption{(a) Real-space map of the total electron density in the presence of 1.3\% disorder consisting of repulsive impurities, $V=0.725$ eV. Total LDOS in the normal state $N(\mathbf r,\omega=0)/N^0(\omega=0)$ (b), self-consistent SC fields $\Delta_{\mu}(\mathbf{r})/\Delta_{\mu}^0$ at $T/T_c^0=1.5$ for $d_{3z^2-r^2}$ (c) and $d_{xz}$ (d). The superscript $0$ denotes parameters of the disorder-free system. $\Delta_{\mu}^0$ refers to the $T=0$ gap value of orbital $\mu$ given by $(0.78,0.78,1.31,0.063,0.055)$ meV in the basis of ($d_{xz}$, $d_{yz}$, $d_{xy}$, $d_{x^2-y^2}$, $d_{3z^2-r^2}$). (e,f) Entropy $S$ (e) and specific heat $C$ (f) versus $T$ comparing the disordered case (colored curves) with the homogeneous system (black curves).}
\label{fig:2}
\end{figure}

Applying conventional wisdom, any sign-changing OP should be quickly destroyed by disorder, and indeed an AG-calculation with e.g. $V=0.725$ eV reveals that merely $\sim0.5\%$ disorder is sufficient to destroy the SC state, as shown in Fig.~\ref{fig:1}(a). This result can be obtained both by a standard T-matrix momentum-based approach\cite{AG1,korshunov,hirschfeld,gastiasoro16} and by a real-space BdG calculation disallowing spatial modulations of density and SC. However, disorder {\it will} induce spatial modulations, and for systems like FeSCs and cuprates, where the coherence length $\xi$ is a few nanometer, AG-theory is no longer applicable. Consider for concreteness a $1.3\%$ disorder concentration producing the nano-scale density modulations shown in Fig.~\ref{fig:2}(a). In Fig.~\ref{fig:2}(b) we plot the total normal state LDOS at $E_F$ $N(\mathbf r)/N^0$, revealing large LDOS enhancements compared to the disorder-free system $N^0$. The formation of these enhancements near the disorder sites can be traced to the generation of resonant states in the off-Fermi-level $e_g$-dominated bands,\cite{gastiasoro13,gastiasoro15} which in turn drive a large enhancement of the SC OP in those orbitals, as seen in Fig.~\ref{fig:2}(c) which shows $\Delta_{e_g}/\Delta^0_{e_g}$  at $T /T_c^0 = 1.5$ i.e. well {\it above} the homogeneous $T_c^0$. Through the coupling to the $t_{2g}$ orbitals dominating the bands near $E_F$, this enhancement leaks into the OP $\Delta_{t_{2g}}$ of these orbitals as seen in Fig.~\ref{fig:2}(d), thereby supporting the entire condensate to stay SC even at $T > T_c^0$.  Therefore, allowing for full freedom in orbital and spatial indices, the resulting SC OP is remarkably robust. The final impurity configuration-averaged $T_c$ is shown by the red curve in Fig.~\ref{fig:1}(a). Here $T_c$ is defined as the highest $T$ where {\it all} sites acquire a finite OP. This definition of $T_c$ marks the onset of a fully connected SC, and is also consistent with the onset of entropy loss and a concomitant step in the specific heat as seen from Fig.~\ref{fig:2}(e,f). 

\begin{figure}[b]
 \begin{center}
 \includegraphics[width=\columnwidth]{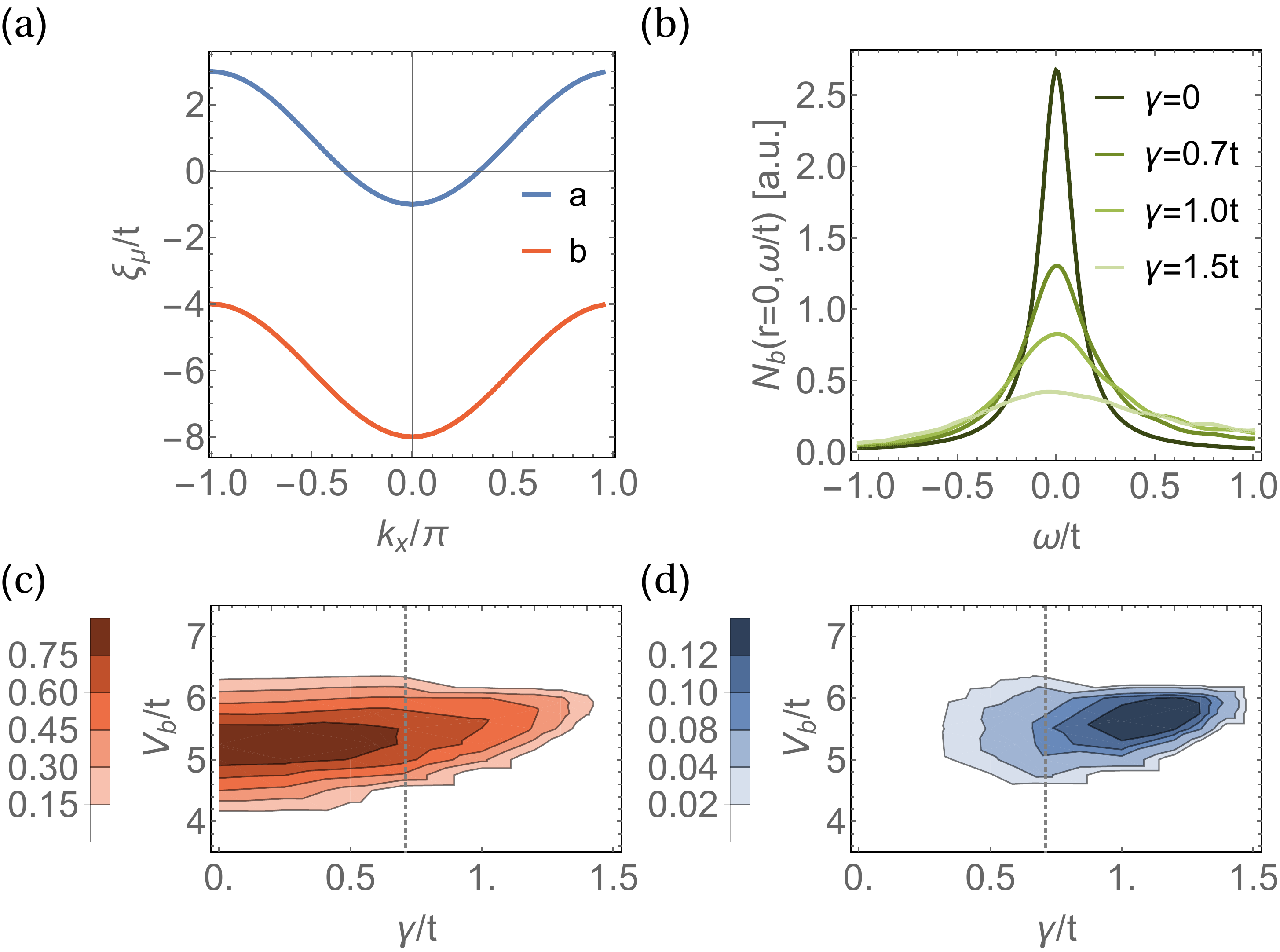}
\end{center}
\caption{(a) Band structure for the two-band toy-model along $\mathbf k=(k_x,\pi/2)$. (b) $N_b(\mathbf r_0,\omega)$ for a resonant potential ($V_b=1/\mbox{Re}[g_{b}^0(\omega=0)]$, $V_a=0$) at the impurity site for different $\gamma$. Note that here we set $V_a=0$ to most clearly demonstrate the origin of the $T_c$-enhancement. For the results in Fig.~\ref{fig:2} we used a more realistic orbital independent potential as stated above. The finite width at $\gamma=0$ arises from an imposed broadening $\eta=T$. (c,d) Self-consistent induced fields $\Delta_{\mu}(\mathbf{r}=0)-\Delta_{\mu}^0$ in units of $t$ for bands $b$ (c) and $a$ (d) as a function of $V_b$ and $\gamma$ for onsite pairing $\Gamma=1.93t$ and $k_BT=0.1t$. The dotted lines in (c,d) show the $\gamma$ above which $\Delta_{\mu}^0=0$ in the disorder-free system at this $T$.}
\label{fig:3}
\end{figure}

{\it Toy model of scenario 1.} In order to illuminate the mechanism for LDOS and $T_c$-enhancements presented above, we analyse a simplified two-band lattice model  
\begin{equation}
\label{eq:Htoy}
\mathcal{H\!}=\!\!\! \sum_{\mu\nu\sigma \mathbf k} [\xi_\mu(\mathbf k)\delta_{\mu\nu} + \gamma\delta_{\mu\overline{\nu}}] \hat{c}^\dagger_{\mathbf k\mu\sigma} \hat{c}_{\mathbf k\nu\sigma} +  \sum_{\mu\mathbf k}  \Delta_\mu\hat{c}^\dagger_{\mathbf k\mu\uparrow}  \hat{c}^\dagger_{-\mathbf k\mu\downarrow} + \mbox{H.c.}
\end{equation}
with dispersion given by $\xi_\mu(\mathbf k)=-2t [\cos(k_x) + \cos(k_y)]-\epsilon_\mu$ with $\epsilon_a=t$ and $\epsilon_b=-6t$, and a coupling $\gamma$ between the two bands, see Fig.~\ref{fig:3}(a). Note that for simplicity for this toy model illustration we include conventional on-site $s$-wave SC as opposed to the NNN pairing of the FeSCs case studied above. The connection to the previous section is that the $a$ ($b$) states dominate the near-Fermi-level (off-Fermi-level) bands corresponding to the $t_{2g}$ ($e_g$) dominated bands in the case of FeSCs. 

In the presence of a pointlike impurity at the site $\mathbf r_0=(0,0)$, the full Green's function is given by
 \begin{equation}
  \mathcal{\hat G}(\mathbf r_i,\mathbf r_i;i\omega_n)\!=\!\mathcal{\hat G}^0(i\omega_n)\!+\!\mathcal{\hat G}^0(\mathbf r_i;i\omega_n)\mathcal{\hat T}(i\omega_n)\mathcal{\hat G}^0(-\mathbf r_i;i\omega_n)
 \end{equation}
where $\mathbf r_i$ denotes the position of the $i$'th lattice site, and $\mathcal{\hat T}(i\omega_n)=\left(\mathbb{I}-\hat V \mathcal{\hat G}^0(i\omega_n)\right)^{-1}\hat V$ and $\mathcal{\hat G}^0(i\omega_n)=\sum_{\mathbf k}\mathcal{\hat G}^0(\mathbf k;i\omega_n)$. In order to expose the $T_c$-enhancement mechanism, let us focus on $T > T_c^0$. In that case, and for a band-diagonal $\hat V$, the impurity states satisfy $\det \left[\mathbb{I}- \hat V\hat G^0(i\omega_n)\right]=\prod_{\mu} \left(1-V_{\mu} g_{\mu}^0(i\omega_n)\right)=0$,
where $\left(g_{\mu}^0(i\omega_n)\right)^{-1}=i\omega_n-\xi_\mu-\gamma^2( i\omega_n-\xi_{\bar\mu})^{-1}$ refers to the local Green's function of band $\mu$ in the homogeneous normal state, and $\bar\mu\neq\mu$.
Therefore, a band with $\mbox{Im}[g_{\mu}^0(\Omega)]\approx0$ exhibits a sharp resonant state at energy $\Omega$ for a potential satisfying $V_{\mu}=1/\mbox{Re}[g_{\mu}^0(\Omega)]$. 
Here, band $b$ is gapped around $E_F$ for small couplings $\gamma$ ($N_b^0\approx0$), and displays a correspondingly sharp resonant state at $E_F$ for $V_b=1/\mbox{Re}[g_{b}^0(\omega=0)]$, as shown in Fig.~\ref{fig:3}(b). As $\gamma$ increases the resonant state broadens due to the finite DOS of the $a$ band near $E_F$, and the LDOS enhancement $N_b$ drops.

The self-consistent gap at the impurity site obtained by solving the associated BdG equations on $40\times 40$ lattices are shown in Fig.~\ref{fig:3}(c,d) for both band $a$ and $b$ as a function of $\gamma$ and $V_b$. As seen from Fig.~\ref{fig:3}(c), the LDOS enhancement of $b$ induces a large corresponding local enhancement of $\Delta_b(\mathbf r_0)-\Delta_b^0$. However, there is no such LDOS increase for band $a$ (not shown), yet $\Delta_a(\mathbf r_0)-\Delta_a^0$ is also significantly enhanced as seen from Fig.~\ref{fig:3}(d). The origin of the increased $\Delta_a(\mathbf r_0)$ is found in the coupling to $\Delta_b(\mathbf r_0)$ as  seen by linearizing the gap equation in the presence of the impurity at $\mathbf r_0$ for small $\gamma$, i.e. $\gamma,\Delta_a(\mathbf r_0)<<\Delta_b(\mathbf r_0)$, obtaining 
\begin{equation}
 \label{eq:Daapprox}
 \Delta_a(\mathbf r_0)\propto V_bF(\Delta_b(\mathbf r_0))\gamma^2,
\end{equation}
where $F(\Delta_b(\mathbf r_0))$ is an increasing function of $\Delta_b(\mathbf r_0)$ that vanishes linearly in the limit $\Delta_b(\mathbf r_0)\rightarrow 0$. Thus, the LDOS enhancement in $b$ directly increases $\Delta_b(\mathbf r_0)$ which indirectly boosts $\Delta_a(\mathbf r_0)$ through the coupling of the bands $\gamma\neq0$ as seen by Eq.~(\ref{eq:Daapprox}). We stress that this result illustrates the main mechanism behind the enhanced superconductivity seen in the FeSC case, see Fig. \ref{fig:2}. Of course, in the realistic FeSC case used above there is a potential in all orbitals, which we excluded for simplicity in the toy model by setting $V_a=0$. In the realistic FeSC case what our calculations show is that the pair breaking in the t$_{2g}$ orbitals (produced by the potential in the  t$_{2g}$ orbitals) is not strong enough the destroy the $T_c$-enhancement when the e$_g$-orbitals are "on resonance". At low $T$, $\Delta_{t_{2g}}$ is indeed suppressed near impurity sites, but the locally boosted $\Delta_{e_{g}}$ is still strong enough to uphold a finite $\Delta_{t_{2g}}$ at all sites in a range of $T$ above $T_c^0$, which is the important finding of our work.

\begin{figure}[b]
 \begin{center}
 \includegraphics[width=\columnwidth]{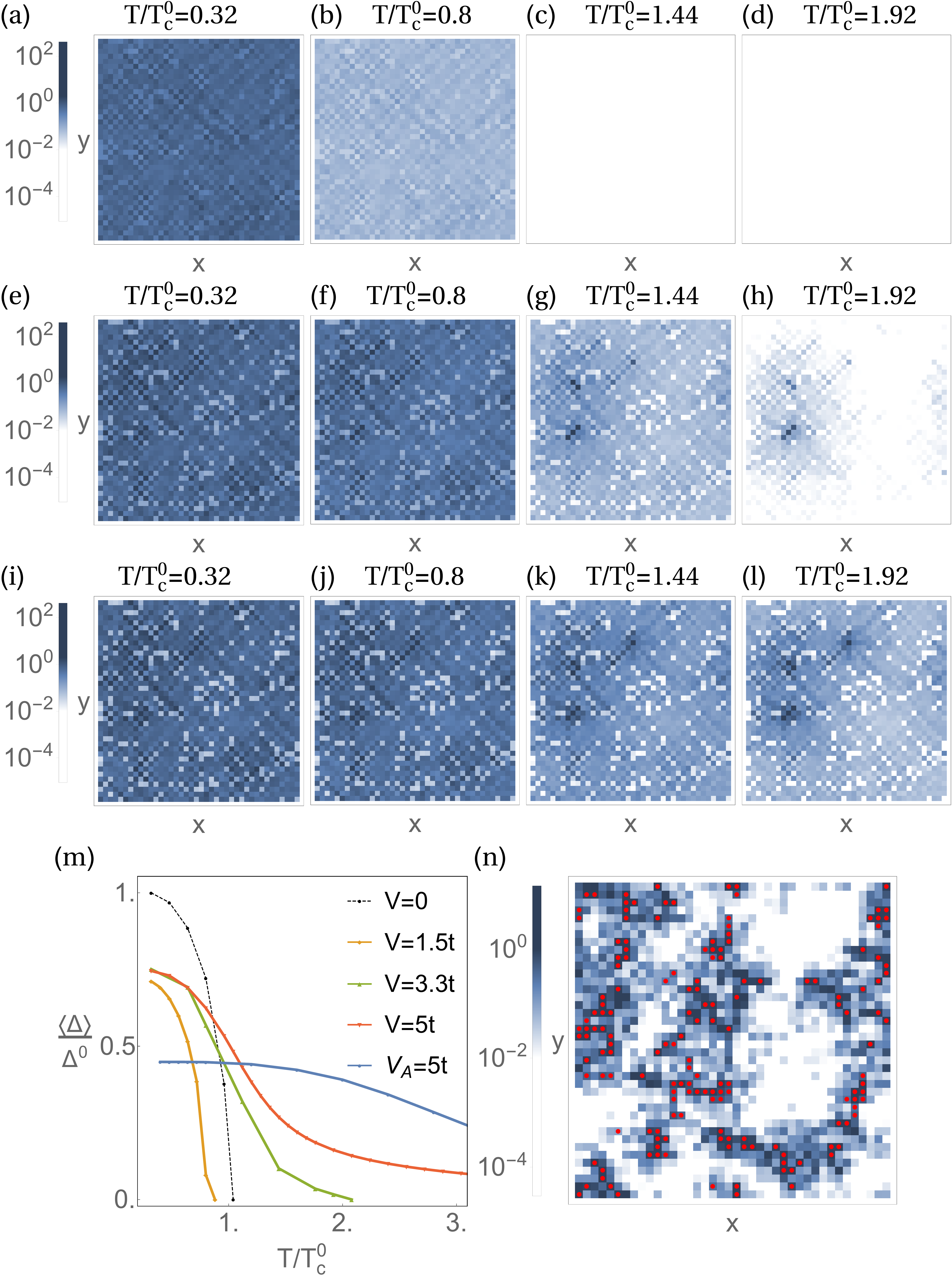}
\end{center}
\caption{(a-l) Real-space maps of $\Delta(\mathbf r)/\Delta^0$ versus $T$ (rows) for a $15\%$ disordered system with varying impurity strength $V=1.5t$ (a-d), $V=3.3t$ (e-h), $V=5t$ (i-l) for a conventional $s$-wave SC in a one-band model. (m) Spatially-averaged SC OP $\langle\Delta\rangle/\Delta^0$ versus $T$ for varying disorder strength $V$, and for Anderson disorder with $V_A \in [-5,5]t$ (blue curve). The clean case is shown by the black curve. (n) Real-space map of $\Delta(\mathbf r)/\Delta^0$ at $T/T_c^0=1.93$ for the case of Anderson disorder. The red dots indicate sites with large $T_c(\mathbf r)/T_c^0 \propto \exp[-1/(|U|N(\mathbf r))]/\exp[-1/(|U|N^0)] > 0.4$, displaying the clear correlation between the local LDOS enhancements and the increased SC.}
\label{fig:4}
\end{figure}

From Fig.~\ref{fig:3}(c,d) it is evident that the $T_c$-enhancement hinges on the effect that resonant states are created near $E_F$, and that the coupling to the Fermi-level-relevant band is finite but weak enough to not destroy the resonant state itself. In materials with properties outside this "golden range", disorder operates as pair-breakers in unconventional SC and lower $T_c$. There may be materials, however, where actual $T_c$-enhancements are not observed but very slow $T_c$-suppression rates are obtained due to the effect described above.\cite{gastiasoro16} For FeSCs, such slow $T_c$-suppression rates have been measured for Ru-substituted LaOFeAs.\cite{satomi,sanna1,sanna2} Finally, we note that interesting $T_c$-enhancements may also be expected in Kondo systems with $T_K>T_c$, since the screened moments produce large LDOS enhancements, in this case guaranteed at $E_F$. The anomalously high $T_c$ observed in the charge-Kondo system Pb$_{1-x}$Tl$_x$Te, where resonant impurity states were recently shown to be crucial for the SC phase, may be related to the scenario presented here.\cite{chargekondo1,dzero,chargekondo2}

{\it Scenario 2.} The above results raise the question of what happens to $T_c$ in conventional SC with sign-preserving OP. For the cases shown in Figs.~\ref{fig:1}(a)-\ref{fig:2}, the answer is that the $T_c$-enhancement is even more pronounced because of the absence of pair-breaking.  However, even for disorder strengths off resonance, a substantial $T_c$-enhancement exists for sign-preserving gap functions with large enough disorder concentrations $n_{imp}$.\cite{Feigelman,Burmistrov,Mayoh} This can be demonstrated by the one-band attractive Hubbard model with NN (NNN) hopping $t$ ($t'=-0.3t$), and filling 0.85, and $s$-wave OP stabilized by onsite attraction $|U|=0.8t$, producing a $\Delta^0=0.022t$ at $T=0$, and $k_BT_c^0=0.0135t$. In Fig.~\ref{fig:4}(a-l) we show the $T$-dependence of $\Delta(\mathbf r)/\Delta^0$ in real-space $40\times 40$ maps for a $15\%$ disordered system with different impurity strengths $V$. As seen, for strong enough $V$ the disorder stabilizes large regions of finite $\Delta(\mathbf r)$ well above $T_c^0$. In Fig.~\ref{fig:4}(m) we show the spatially-averaged $\Delta(\mathbf r)$ as a function of $T$, clearly demonstrating the enhanced SC for the disordered case.  We have also studied Anderson disorder, and found similar behavior, shown by the blue curve in Fig.~\ref{fig:4}(m). The origin of the $T_c$ enhancement is favorable centers of enhanced LDOS as seen from Fig.~\ref{fig:4}(n) showing the strong correlation between $\Delta(\mathbf r)$ and $N(\mathbf r)$.

Unlike scenario 1, for the results in Fig.~\ref{fig:4}, the high level of disorder prevents one from defining $T_c$ in terms of simple steps or discontinuities in thermodynamic quantities. The spatially averaged $\langle\Delta\rangle$ also does not easily allow for a sound definition of $T_c$ because the OP breaks up into disconnected regions at large $T$. Therefore, for this case we define $T_c$ as the highest $T$ where all edges of the system are fully connected by gap amplitudes of at least $p\%$ of $\Delta^0$, the $T=0$ homogeneous OP. Fig.~\ref{fig:1}(b) shows the resulting $T_c/T_c^0$ curves for different $p$ thresholds, with a substantial upturn for strong disorder. Obviously the value of $p$ affects the magnitude of the $T_c$-enhancement, but not the existence of a disorder-generated $T_c$-enhancement itself. 

For the cases shown in Fig.~\ref{fig:4}, one may estimate the mean free path $\l=v_F \tau$ from the scattering rate $\tau^{-1}=2 \pi n_{imp} N^0 V^2/(1+[N^0V]^2)$, yielding that for the cases with $V \gtrsim(2-3)t$, $\l \sim 1-2$ lattice spacings. Therefore these cases are in the Anderson localized limit, and the existence of a $T_c$-enhancement is consistent with the findings by Burmistrov {\it et al.}\cite{Burmistrov} We ascribe the reason that $T_c$-enhancements were not previously seen in numerical simulations to the small system size and very large SC OP used in those studies.\cite{trivedi1996,ghosal_phase,bouadim}

It remains interesting to extend the current studies to include phase fluctuations, potentially important for inhomogeneous systems with regions of low superfluid density.\cite{larkin} In general, phase fluctuations lower the mean-field $T_c$, but the reduction depends strongly on dimensionality and the spatial structure of the modulations driving the inhomogeneity. For 3D systems, and when $\xi$ is of the same scale as the disorder-generated density modulations, the mean-field $T_c$ is not expected to be strongly affected by phase fluctuations.\cite{Burmistrov,martin,larkin} 

We have studied mean-field $T_c$-enhancements in both conventional and unconventional superconductors from disordering with nonmagnetic impurities. Our results suggest a path to engineer systems with larger $T_c$ by introducing suitable amounts of disorder. We focussed on superconductivity, but similar effects may be also expected for systems with other preferred symmetry breaking.

We thank I. S. Burmistrov, P. J. Hirschfeld, A. Kreisel, A. T. R\o mer, and Avraham Klein for useful discussions. We acknowledge support from a Lundbeckfond fellowship
(Grant A9318).

\pagebreak
\widetext
\begin{center}
\textbf{\large Supplemental Material: "Enhancing Superconductivity by Disorder"}
\end{center}
\setcounter{equation}{0}
\setcounter{figure}{0}
\setcounter{table}{0}
\setcounter{page}{1}
\makeatletter
\renewcommand{\theequation}{S\arabic{equation}}
\renewcommand{\thefigure}{S\arabic{figure}}
\renewcommand{\bibnumfmt}[1]{[S#1]}
\renewcommand{\citenumfont}[1]{S#1}

Here, we provide computational details of the results presented in the first part of the paper, relevant to multi-band systems (scenario one).

\section{Model and Selfconsistency equations}


The starting Hamiltonian defined on a two-dimensional lattice is given by
\begin{equation}
 \mathcal{H}=\mathcal{H}_{0}+\mathcal{H}_{BCS}+\mathcal{H}_{imp},
\end{equation}
describes a superconducting system in the presence of disorder.
We use a five-orbital tight-binding band relevant to the 1111 pnictides~\cite{Sikeda10} 
\begin{equation}
\mathcal{H}_{0}=\sum_{\mathbf{ij},\mu\nu,\sigma}t_{\mathbf{ij}}^{\mu\nu}\hat c_{\mathbf{i}\mu\sigma}^{\dagger}\hat c_{\mathbf{j}\nu\sigma}-\mu_0\sum_{\mathbf{i}\mu\sigma}\hat n_{\mathbf{i}\mu\sigma}.
\end{equation}
We stress that for the 1111 systems a two-dimensional model should be appropriate since the dispersion along the $k_z$ direction is essentially absent.\cite{Sikeda10}  
The chemical potential $\mu_0$ adjusts the average electron density to 6.0 electrons per site. 
Superconductivity is included by a BCS-like term
\begin{equation}
 \mathcal{H}_{BCS}=-\sum_{\mathbf{i}\neq \mathbf{j},\mu\nu}[\Delta_{\mathbf{ij}}^{\mu\nu}\hat c_{\mathbf{i}\mu\uparrow}^{\dagger}\hat c_{\mathbf{j}\nu\downarrow}^{\dagger}+\mbox{H.c.}],
\end{equation}
with $\Delta_{\mathbf{ij}}^{\mu\nu}=\sum_{\alpha\beta}\Gamma_{\mu\alpha}^{\beta\nu}(\mathbf{r_{ij}})\langle\hat{c}_{\mathbf{j}\beta\downarrow}\hat{c}_{\mathbf{i}\alpha\uparrow}\rangle$ being the superconducting order parameter, and $\Gamma_{\mu\alpha}^{\beta\nu}(\mathbf{r_{ij}})$ denoting the effective pairing strength between sites (orbitals) $\mathbf{i}$ and $\mathbf{j}$ ($\mu$, $\nu$, $\alpha$ and $\beta$). 
In agreement with a general $s^\pm$ pairing state, we include next-nearest neighbor (NNN) intra-orbital pairing, $\Gamma_{\mu}\equiv\Gamma_{\mu\mu}^{\mu\mu}(\mathbf{r_{nnn}})=0.208$ eV.
The last term in the Hamiltonian introduces non-magnetic disorder in the system,
\begin{equation}
\mathcal{H}_{imp}=\sum_{\{\mathbf{i^*}\}\mu\sigma}V_{\mu} \hat c_{\mathbf{i^*}\mu\sigma}^{\dagger}\hat c_{\mathbf{i^*}\mu\sigma}
\end{equation}
Here $V_{\mu}$ denotes the impurity potential in orbital $\mu$ at the disorder sites given by the set $\{\mathbf{i^*}\}$ coupled to the charge density of the itinerant electrons. This potential is assumed to be orbital independent, in overall agreement with DFT findings.

By using the spin-generalized Bogoliubov transformation,
\begin{align}
\label{eq:bogtrans}
\hat{c}_{\mathbf{i}\mu\sigma}&=\sum_{n}(u_{\mathbf{i}\mu\sigma}^{n}\hat{\gamma}_{n\sigma}+v_{\mathbf{i}\mu\sigma}^{n*}\hat{\gamma}_{n\overline{\sigma}}^{\dagger}),
\end{align}
we arrive to the Bogoliubov-de Gennes (BdG) equations 
\begin{eqnarray}
\begin{pmatrix}
\hat{\xi}_{\uparrow} & \hat{\Delta}_{\mathbf{ij}}\\
\hat{\Delta}_{\mathbf{ji}}^{*} & -\hat{\xi}_{\downarrow}^{*} 
\end{pmatrix}
\begin{pmatrix}
 u^{n} \\ v^{n} 
\end{pmatrix}=E_{n}
\begin{pmatrix}
 u^{n} \\ v^{n} 
\end{pmatrix}.
\end{eqnarray}
The transformation
$\begin{pmatrix}
   u_{\uparrow}^{n} && v_{\downarrow}^{n} && E_{n\uparrow}
 \end{pmatrix}
\rightarrow
\begin{pmatrix}
 v_{\uparrow}^{n*} && u_{\downarrow}^{n*}&&-E_{n\downarrow}
\end{pmatrix}$
maps two of the equations onto the other two and thus we drop the spin index from the eigenvectors and eigenstates of the BdG equations.
The matrix operators are defined as:
\begin{align}
 \hat{\xi}_{\sigma}u_{\mathbf{i}\mu}&=\sum_{\mathbf j\nu}\left[t_{\mathbf{ij}}^{\mu\nu}+\delta_{\mathbf j \mathbf i}\delta_{\mu\nu}\left(-\mu_0+V_{\mu}\delta_{\mathbf{i\{i^*\}}}\right) \right] u_{\mathbf{j}\nu},\\
\hat{\Delta}_{\mathbf{ij}}^{\mu\nu}u_{\mathbf{i}\mu}&=-\sum_{\mathbf{j}\nu}\Delta_{\mathbf{ij}}^{\mu\nu}u_{\mathbf{j}\nu},
\end{align}
where the notation $\delta_{\mathbf{i\{i^*\}}}$ refers to the fact that the $V_{\mu}$-term is only present if $\mathbf{i} \in \{\mathbf{i^*}\}$.
The five-orbital BdG equations are solved on $30\times30$ lattices with stable solutions found for a given temperature T through iterations of the following self-consistency equations 
\begin{align}
\label{eq:bdg}
  n_{\mathbf{i}\mu\uparrow} &= n_{\mathbf{i}\mu\downarrow}=\sum_{n}|u_{\mathbf{i}\mu}^{n}|^{2}f(E_{n}),\\\nonumber
\Delta_{\mathbf{ij}}^{\mu}&=\Gamma_{\mu}\sum_{n}u_{\mathbf{i}\mu}^{n}v_{\mathbf{j}\nu}^{n*}f(E_{n}),
\end{align} 
where $\sum_n$ denotes summation over all eigenstates $n$, and $f(E_n)$ is the Fermi distribution function. The chemical potential $\mu_0$ is adjusted in each case to maintain the average density of the system, $n_0=\frac{1}{\mathcal{N}}\sum_{i\mu}\left( n_{\mathbf{i}\mu\uparrow}+ n_{\mathbf{i}\mu\downarrow}\right)=6.0$. We stress that the solutions are fully unrestricted and allowed to vary on all lattice sites and orbitals.
The superconducting order parameter shown in the main manuscript is the bond averaged singlet component:
\begin{equation}
 \Delta_{\mathbf i\mu}=\frac{1}{4}\sum_{\mathbf j}\frac{1}{2}(\Delta_{\mathbf{ij}}^{\mu}+\Delta_{\mathbf{ji}}^{\mu})
\end{equation}
where $\mathbf j$ are four next nearest neighbors.
The inclusion of several impurities leads to a spatially varying order parameter $\Delta_{\mathbf{i}\mu}$.
We define $T_c$ as the highest temperature where all sites acquire a finite $\Delta_{\mathbf{i}\mu}$.



\end{document}